# Assessing public-private research collaboration: is it possible to compare university performance?[1]

*Giovanni Abramo*[a,b,*], *Ciriaco Andrea D'Angelo*[a] *and Marco Solazzi*[a]

[a] Laboratory for Studies of Research and Technology Transfer
School of Engineering, Dept of Management
University of Rome "Tor Vergata"

[b] National Research Council of Italy

**Abstract**

It is widely recognized that collaboration between the public and private research sectors should be stimulated and supported, as a means of favoring innovation and regional development. This work takes a bibliometric approach, based on co-authorship of scientific publications, to propose a model for comparative measurement of the performance of public research institutions in collaboration with the domestic industry collaboration with the private sector. The model relies on an identification and disambiguation algorithm developed by the authors to link each publication to its real authors. An example of application of the model is given, for the case of the academic system and private enterprises in Italy. The study demonstrates that for each scientific discipline and each national administrative region, it is possible to measure the performance of individual universities in both intra-regional and extra-regional collaboration, normalized with respect to advantages of location. Such results may be useful in informing regional policies and merit-based public funding of research organizations.

---

# 1. Introduction

In the current context of the global knowledge-based economy, the governments of industrialized nations perceive the necessity of strengthening domestic scientific infrastructures. Special attention is devoted to public research institutions (PRIs), meaning universities and publicly funded research institutes. The efforts focus on two main objectives: i) increased productivity from research activities, and ii) more efficient public to private technological transfer. As part of their mix of expenditures and measures, a number of governments now operate national research assessment exercises. Although the practice of each country is not identical, the objectives of these exercises are generally to: i) stimulate greater scientific productivity through publicizing the performance of PRIs; ii) improve the allocative efficiency of public funding for research, by distributing a portion of funds on the grounds of merit; iii) reduce information asymmetry on research quality between supply (PRIs) and demand (private enterprises, students, etc.), thus achieving greater market efficiency. There is also increasing awareness, at the political and administrative levels, of the need for more proactive economic roles for PRIs. PRIs respond by devoting more attention to technological transfer, thus contributing to regional economic development and competitiveness, while still striving to maintain their traditional grand roles of research and education. Universities thus become ever more active in development of regional innovation systems (Gunasekara, 2006), in what is now called a "third role", in step with their more traditional understood functions (Etzkowitz and Leydesdorff, 1997).

With varying emphasis and results, the individual PRIs of each nation are now striving to reorient their missions, strategies, management systems and practices to enable fulfillment of their expected new role in development. This now brings about a necessity to monitor and evaluate, not only the research activity of PRIs, but also their success in achieving technological transfer. Evaluation of success in technology transfer becomes an important consideration from the point of view of i) management of PRIs, as part of strategic control; ii) policy-makers, organizing national research assessment exercises, and iii) management and policy-makers, examining the impact of policies to stimulate technological transfer. This type of evaluation first requires identification of all the methods through which technological transfer can occur (for a literature review on the issue see Bozeman, 2000). When it comes to the actual comparative evaluations conducted to date, the forms of technology transfer most commonly examined are licensing and, more recently, creation of spin-off companies. These are two important forms of public-private transfer, but the focus on them is partially motivated by the relative ease with which results can be measured. Where ad-hoc monitoring systems are put in place (see for example Pressman et al., 1995), the intensity and impact of these forms of technology transfer can be measured within individual research institutions. Comparative measurement is also possible at a broader national level, depending to some extent on the reliability and exhaustiveness of national databases (AUTM, 2008; Abramo, 2006).

Public-private research collaboration is another important channel through which PRIs transfer knowledge to industry. The main reasons for PRI researchers to collaborate with enterprises are as follows: i) to obtain resources that would not otherwise be available (financing, equipment and facilities); ii) to experiment with their concepts in the "real" world (translate theoretical knowledge into practical applications); iii) to exchange knowledge with others (Beaver, 2001; Lee, 2000; Meyer-Krahmer and Schmoch, 1998). Private companies are motivated to engage in research collaborations with PRIs in order to tap their potential to generate new knowledge, and are generally willing to contribute with financial resources. Since the 1970s public-private research collaboration has attracted ever greater interest (Caloghirou et al., 2001), to the point of now becoming a central theme of national and supra-national policies (e.g. European Union). This is in spite of the fact that difficulties are recognized, and the known argument that collaboration does not always lead to innovation (Link and Tassey, 1989). Indeed, there are great differences in organizational structure, mission and conception of research between the two sides, and difficulties in managing the results from actual collaboration. While PRI researchers are eager to publish the results of their research activities, private companies are concerned that knowledge spillovers may benefit competitors. Cohen et al. (1998) have described some of these difficulties using the term of "uneasy partners". Still, there is a general opinion that polices stimulating research collaboration are useful and necessary, a view that is strengthened by observations that private enterprises increasingly resort to cooperation with universities (OECD, 2007) and that university researchers who collaborate more with industry demonstrate a higher level of productivity with respect to their colleagues[2]. Given that technological transfer is taking on a more important role in the mission of PRIs and in national and regional development policies, and that public-private research collaboration is one of the most effective methods to achieve such transfer, the measurement of such collaborations is certainly called for. However, the measurement of comparative performance of PRIs in intensity of collaboration results as difficult, because there is a lack of available data. Timely measurement of the impact of collaboration, to enable subsequent management and policy decisions, results as still more difficult.

In this work, the authors offer a first model to assist in comparative measurement of PRI performance in technology transfer, for the dimension of research collaboration with industry. The potential and limits of the measurement system are illustrated by applying it to the evaluation of the entire Italian university system ("hard science" disciplines only) in the period 2001-2003. The intent is not to provide a ranking of university performance, rather to present the model. The application of the model, the authors hope, should be useful for all interested stakeholders. University management may use it to compare the performance of their departments against similar departments in other universities, in terms of their ability to collaborate with industry. National and regional policy-makers may use the system to evaluate the effectiveness of their policies and interventions in favor of research collaboration. Local policy makers may use it to monitor the extent and benefit of knowledge transfer generated through collaboration by universities in their region, while at the national level, policy-makers can use the system to support strategies for harmonizing and integrating regional interventions. Finally, in keeping with the desired third role of universities, the system can also be useful in upgrading current national research assessment exercises, because it permits the additional element of examining results in technology transfer, for the dimension of research collaboration with industry.

The next section of this paper gives a first presentation of the model, together with the methodological approach and dataset for the work. The third section describes the general character of all Italian regions, in terms of their university-industry research collaboration activity, at intra and extra-regional levels. The fourth section presents an example of how the model can be applied to the analysis of the details of university-industry collaborations, by scientific sector, and the comparative measurement of university performance. The last section provides conclusions and indicates further potential directions for development and application of the measurement system.

---

[2] This result is seen in studies from various nations including examples from Australia (Harman, 1999), Norway (Gulbrandsen and Smeby, 2005) and Italy (Abramo et al., 2009).

## 2. Model, data and methods

### 2.1 *Model*

While PRI-industry research collaboration cannot be considered a one-way transfer of knowledge, the partners' underlying motivations in activating a collaboration lead us to assume that the balance of the knowledge transfer is in favor of industry. In most cases in fact it is the private company that contributes with financial resources to the joint research project. Consequently, we refer to PRIs as "supply" and to private companies as "demand" in the market for research collaboration. A general lack of homogenous and complete databases on public-private research collaborations presents obvious difficulties in developing instruments for evaluating PRI performance in this area. The entire issue of research and development is somewhat sensitive, and for strategic reasons, private partners may be particularly resistant to releasing data.

The approach proposed here, and applied to the case of Italian universities, is to identify collaborations through the proxy of scientific articles published in co-authorship between universities and industry, as listed in bibliometric databases such as Web of Science, Scopus, etc. The limits of using a bibliometric approach have been noted in literature (Melin and Persson, 1996; Katz and Martin, 1997; Laudel, 2002), with particular emphasis on the facts that scientific collaborations do not always lead to publications, and that real collaboration does not always exist among scientists listed as co-authors. However, we find no reasons why the distribution of complementary cases should not be homogeneous among universities. The eagerness of academics to publish should make co-authored publications a representative proxy of all research collaborations, although a relevant survey has not been conducted by the authors to test the proxy. The bibliometric method has numerous advantages, such as its non-invasive character, relatively low cost, and the homogenous and exhaustive character of the data. Meanwhile, in the absence of databases specifically on research collaborations, alternative measurement methods present other types of weakness. Traditional surveys, with information collected from the subjects, generally present problems of completeness and thus of reliability in inferential analyses. An analysis of patent applications filed jointly by private companies and universities would generally provide a useful complement, but in Italy the number of patents filed by all universities is notoriously low. Using the Espacenet search engine we have identified 284 patents filed by Italian universities between 2001 and 2003[3]. Of these, only 20 were co-owned with private companies. Because co-ownership does not necessarily indicate the actual authorship of a patent, the number of university patents stemming from true collaboration with industry may actually be less than 20. On the other hand, many university researchers who file patents would also wish to publish scientific articles on the subject, therefore the analysis of publications could capture some information concerning patents. For these reasons we have chosen not to include patents in our analysis.

---
[3] Legislation in 2001 introduced the so called "academic privilege", presumably resulting in additional patents filed by university researchers, but relevant data are not readily available, making the identification of joint patents very difficult.

The model has been applied to the assessment of Italian university performance in research collaboration with domestic industry. We have excluded collaborations with international companies from the analysis because our intent is to present a model suitable for comparing university contributions to regional and national economic development, through collaboration with local and domestic industry. From this perspective, it is in the regional and national governments' interest that it is their own respective territories, rather than those of "competitors", which receive the economic benefits of university research. As a consequence, university collaborations with foreign companies, if the knowledge transfer balance is indeed in favor of industry, might actually be counter-productive[4]. On the other hand, international collaboration is an indicator of prestige of a research group. The measurement of it would be more appropriate when assessing research excellence, rather than performance in technology transfer.

The model places particular attention on describing the phenomenon of research collaboration with respect to two significant variables: the scientific sector concerned and the location of the partners. This focus is intended to provide detailed information, useful to decision-makers. But it is also a necessity in order to avoid the inevitable distortions that would result from aggregate-level analysis. Aggregate analysis fails to respect the differences that occur between scientific sectors, both in intensity of publication and in collaboration (Abramo et al., 2008a), and also among regions for a particular sector, as a function of differing concentrations of private enterprise and stimulus policies adopted by local governments (Abramo et al., 2009). It has also been observed that the knowledge spillovers vary between industrial sectors (Anselin et al., 2000), and that spillovers are favored between regions that have similar characteristics in terms of scientific-technological levels and sectorial emphasis (Maggioni and Uberti, 2005). The importance of geographic proximity between producers of new knowledge (PRIs) and users (private enterprises) is also known. Knowledge spillovers are primarily local in their characteristics (Anselin et al., 1997; Jaffe, 1989; Autant-Bernard, 2001; Parente and Petrone, 2006), with the importance of local proximity resulting from the actual mechanisms for transfer of knowledge, which in turn frequently depend on face-to-face interactions and mobility of personnel (Kaufmann and Tödtling, 2001). Regions can also adopt different policies and incentives for university-industry collaboration, which can then also vary by sector.

Further detailed analyses able to weigh research collaborations in terms of cost and duration have thus far been impeded by the lack of relevant data.

*2.2 Data and methods*

The database used in this particular work is the CD-rom version of the Science Citation Index, SCI[TM], produced by Thomson Reuters. As a first step in the current measurement system, the SCI[TM] was used to extract publications (articles and reviews), published in the period 2001-2003, authored by Italian university scientists. From this initial group of 52,634 publications, involving about 30,000 university authors, the next step was to select those articles co-authored with private companies located in Italy, to arrive at the subset of publications used for further analysis. To do this, we had to identify and reconcile the different ways in which the same research institution was reported in the SCI$^{TM}$ "address" field for the articles. Finally, through a "disambiguation" algorithm, we were able to accurately attribute each publication (with an error below 3%) to its respective academic authors (see Abramo et al., 2008b for details).

---

[4] The U.S. Bay-Dohle Act provides an example of a measure to protect domestic industry by regulating university licensing: the model is emulated by a number of other countries.

The Italian university system requires that every university researcher be affiliated with a single Italian university and with one specifically named scientific disciplinary sector (SDS)[5], which is in turn part of a broader scientific discipline. It was thus possible, as part of the methodology, to link each publication to the SDS of the pertinent university authors.

The field of observation for the analysis consists of the "hard science" disciplines: mathematics and computer sciences, physics, chemistry, earth sciences, biology, medicine, agricultural and veterinary sciences, and industrial and information engineering[6], which included a total of 183 SDS. The dataset thus results as the subset of SCI$^{TM}$-listed publications for 2001 to 2003, jointly authored by at least one private enterprise located in Italy and an academic scientist falling in one of these 183 SDS. This includes 1,534 publications (around 3% of total[7]), produced by 58 universities (out of a total of 78 in Italy at the time of analysis) and 438 enterprises.

Based on this dataset, the collaborations are studied at two distinct levels of analysis: one at the level of individual organizations (university-enterprise level) and the other, more refined, at the scientific sector level (SDS-enterprise level).

By "university-enterprise collaboration", we mean a research collaboration between a university and an enterprise that has resulted in exactly one co-authored publication of the dataset under consideration. A generic publication by *m* universities and *n* enterprises corresponds therefore to *m*n* university-enterprise collaborations. The 1,534 co-authored publications resulted in 1,983 university-enterprise collaborations.

For the analysis at the sectorial level it was obviously necessary to identify the scientific sectors involved in each research collaboration. This was assumed to be the SDS to which the university authors belong. This could result in incorrect classification of articles where the university author publishes on an issue outside his official SDS. However, it is reasonable to assume that a university researcher will actually carry out research in his or her own area of specialization, especially if the research involves collaboration with other parties.

In formal terms, for a generic publication *i* of the dataset, one has:

$$(\text{Number of collaborations at SDS-company level})_i = \left[ \sum_{k=1}^{m_i} \text{SDS\_coauthors}^i_k \right] * n_i$$

where:

---

[5] A list of all SDS may be found at http://www.miur.it/atti/2000/alladm001004_01.htm
[6] The discipline of civil engineering produces few publications listed in the SCI$^{TM}$. For this reason it is not included in the field of analysis.
[7] In the analysis of Hicks and Hamilton (1999), concerning publications by US authors between 1981 and 1994, as indexed in the Science Citation Index, university-industry co-authored publications result as 5.5% of total university publications.

- $m_i$ = number of universities with at least one scientist co-author of publication *i*;
- $SDS\_coauthors^i_k$ = number of SDS, of university *k*, with at least one scientist co-author of publication *i*;
- $n_i$ = number of enterprise co-authors of publication *i*.

As an example of this working definition, an article co-authored by three enterprises and five academic scientists from a single university but from two different SDS would result in six SDS-enterprise collaborations. We would also see six SDS-enterprise collaborations for the case of an article authored by three enterprises and two university scientists in the same SDS but affiliated to different universities. The number of SDS-enterprise collaborations thus takes account of both the number of SDS to which the university authors belong (supply of knowledge) and the number of companies involved (demand). The application of the formula shows the occurrence of 2,363 SDS-enterprise collaborations, in 141 SDS.

The location of each university and enterprise was identified. The comparative evaluation of university performance could thus be conducted at the regional level, thereby taking into account the effects of geographic proximity and regional policies in favor of collaboration, and avoiding any potential distortions. This level of analysis also permitted evaluation of the degree of correspondence between demand and supply of collaboration in each region, and the identification of collaborations between research partners in the same region (intra-regional collaboration) and with partners in other regions (extra-regional collaborations).

Compared to the literature, the methodology used here most closely resembles that of a study by Tijssen et al. (2009)[8], in which a bibliometric approach based on co-authorship data (source: WoS) is used to compare the numbers of university-industry co-authored papers produced by the 350 largest research universities in the world. The study measures the percentage of university-industry co-authored papers in overall scientific production for all the research institutions considered. It distinguishes each university-industry co-authored paper as 'domestic', 'foreign' or both, depending on the nationalities of the research partners involved. The model we present in the current work again uses the research partners' geographic location to apply a distinction (regional and national), but instead of simply counting the co-authored papers it also takes into account the number and typology of partners (university-enterprise levels), and the scientific sector involved in the relationship (SDS-enterprise level). It measures the collaboration performance of each Italian university by scientific discipline, normalizing the performance with respect to the concentration of private enterprises in the university's region.

## 3. Aggregate analysis of collaborations: the context at the regional level

Before analyzing the performance of single universities and their disciplinary sectors it is useful to first provide some brief background information about the Italy's university and industrial systems. Italian R&D expenditures are around 1.1% of GNP, equally split between public and private sectors. Universities are almost totally public, with government funding representing an average of about 85% of their revenues. Italian industry has experienced progressive hi-tech de-specialization in recent years (Gallino, 2003) and is now generally middle and middle-low tech. The composition of the industrial system is also characterized by a disproportionate ratio of small and micro companies. The empirical evidence from technology transfer studies thus far conducted for the Italian case shows a certain situation: i) a strong tendency by the Italian public researcher, compared to his/her counterpart in other countries, to have recourse to publications rather than patents; and ii) a low intensity of public-private technology transfer (Abramo and D'Angelo, 2009).

---

[8] The authors note that the work by Tijssen did not inspire the current work, since it came to their awareness only at the moment that the current work was submitted for publication.

In this section we present a general description of the market for research collaboration in each region, describing the character of supply and demand. This gives a context for the subsequent analysis, showing the differences in terms of numbers of universities and enterprises active in collaboration and their tendency for intra and extra-regional collaboration. The subsequent evaluation of individual universities actually requires this first consideration of their regional context, because an enterprise's choice of its university partners is significantly affected by factors of geographic proximity (Lee and Mansfield, 1996). Universities that are situated close to dynamic and innovative enterprises will generally have an advantage, especially if there is a lack of nearby "competitor" universities[9]. Azagra-Caro (2007) also show that if a region has a low absorptive capability[10], the types of university researchers that are generally more involved in collaboration will then tend to work with relatively large sized enterprises that are situated outside their home region. Regional administrations also adopt different policies and instruments that provide varying incentives for public-private collaboration. Thus the necessary first analysis is at the regional level. But it should also be noted that a particular university could actually be closer to enterprises situated in another administrative region and, other factors being equal, geographic proximity would then favour "extra-regional" collaboration.

The regional distribution of the total 1,983 university-enterprise collaborations in the dataset is presented in Table 1. The regions indicated are the first-level administrative divisions of the Italian nation. The region of Val d'Aosta is not included because the enterprises and the one university situated there did not participate in any collaboration that resulted in a publication: all subsequent analysis is thus limited to the remaining 19 regions. The table shows that there is no industrial demand for collaboration in Basilicata, indicated by the absence of any enterprise that participated in co-authorship collaboration with a national university during the triennium. Meanwhile, there is no supply of collaboration in Molise, as seen by the lack of participation by any university in co-authorship with enterprises.

Lombardy appears in first place in almost all classifications: for number of universities involved in collaborations (nine); number of enterprises that collaborate with universities (a full 166 of the 509[11] registered at the national level); for university-

---

[9] It should be noted that in addition to universities (research institutes) also contribute to the production of new knowledge, but are not fully considered in this work. The current work is primarily intended to describe a measurement system and provide an example of its application to the Italian case: the results should be interpreted in this sense.

[10] From Azagra-Caro (2007): "We follow Cohen and Levinthal's (1990) definition of absorptive capacity: *a limit to the rate or quantity of scientific or technological information that a firm can absorb*. To justify the extension of the concept of absorptive capacity from firms to regions see Niosi and Bellon (2002)".

[11] To enable comparison between regions, each seat of a private enterprise, if it was involved in co-authorship of an article, was considered as a distinct enterprise. For this reason the number of

enterprise collaborations realized by local universities (403) and also by local enterprises (796); and for the percentage of university collaborations with intra-regional enterprises with respect to the total collaborations carried out by its universities (57.82%). Lombardy is a region with a high concentration of private enterprise, and it would be in part because of this that the universities of the region achieved the highest number of collaborations with local enterprises.

Following Lombardy, the regions with the next highest levels of university collaborations with Italian enterprises (column three) are Emilia Romagna (298) and Tuscany (213). For collaborations by enterprises with universities (column six), Lombardy is followed by Lazio (289) and Tuscany (215).

The table is interesting in illustrating the extent to which universities and enterprises establish intra-regional compared to extra-regional collaborations. There are only 2 regions in which the universities are involved more in intra-regional than in extra-regional collaborations (column nine): Lombardy (57.82%) and Sicily (52.54%). This observation should be a cause for regional governments to reflect on how to increase local benefits from new knowledge produced by universities in the region. The high ranking of Lombardy is explained by the notable concentration of industry in the region, and thus of the private research carried out in the region. In fact, approximately 30% of Italy's total private sector research expenditures are concentrated here, representing the highest regional level in the country (ISTAT, 2008). The high percentage of intra-regional collaborations by universities in Sicily could be explained by the fact that it is an island, at the southern tip of the peninsula, making for difficult connections to the rest of the nation. Meanwhile, the lowest levels for intra-regional collaborations by universities are seen in Apulia (6.25%), Umbria (5.08%) and Calabria (0%). These three regions also have the lowest numbers of enterprises involved in collaborations (four for Apulia and Umbria, two for Calabria), meaning that the universities desiring collaboration are forced to target enterprises outside their region.

Examining the counterpart tendencies for behavior of enterprises (column 11), it results that for those in the Marche, Apulia and Umbria, 75% of their collaborations are within their respective regions, meaning that the enterprises choose local universities in three out foru cases. Enterprises in Lombardy, Lazio, and Trentino Alto Adige have the lowest proportion of intra-regional collaboration (respectively 30.30%, 21.80% and 20.00%).

Of the national total of 1,983 collaborations, 690 (34.79%) involved partners within a single region. Comparison of columns nine and eleven shows that the universities tend to collaborate at an intra-regional level (42.35% of collaborations) slightly more than do the enterprises (39.34%). However the difference seen is modest and arises mostly from data for the four regions that are most active in collaboration, Lazio, Lombardy, Piedmont and Tuscany, which are the only regions where the share of collaborations at regional level is higher for universities than for enterprises.

---

"enterprises" increases from 483 to 509.

| (1) Region | (2) Number of universities[i] | (3) Collaborations by universities (percent of total in brackets) | (4) Rank | (5) Number of enterprises[ii] | (6) Collaborations by enterprises (percent of total in brackets) | (7) Rank | (8) Number of intra-regional collaborations | (9) Intra-regional collaborations as percentage (%) of the total university collaborations | (10) Rank | (11) Intra-regional collaborations as percentage (%) of the total enterprise collaborations | (12) Rank |
|---|---|---|---|---|---|---|---|---|---|---|---|
| Abruzzo | 3 | 57 (2.87) | 12 | 9 | 23 (1.16) | 8 | 13 | 22.81 | 9 | 56.52 | 7 |
| Apulia | 4 | 48 (2.42) | 13 | 4 | 4 (0.20) | 15 | 3 | 6.25 | 15 | 75.00 | 1 |
| Basilicata | 1 | 6 (0.30) | 18 | 0 | 0 (0.00) | 19 | NA | NA | NA | NA | NA |
| Calabria | 3 | 13 (0.66) | 17 | 2 | 2 (0.10) | 17 | 0 | 0.00 | 17 | 0 | 17 |
| Campania | 4 | 103 (5.19) | 8 | 11 | 21 (1.06) | 10 | 13 | 12.62 | 11 | 61.90 | 6 |
| Emilia Romagna | 4 | 298 (15.03) | 2 | 61 | 196 (9.88) | 4 | 93 | 31.21 | 6 | 47.45 | 8 |
| Friuli VG | 3 | 60 (3.03) | 9 | 8 | 21 (1.06) | 10 | 15 | 25.00 | 8 | 71.43 | 5 |
| Lazio | 6 | 160 (8.07) | 5 | 64 | 289 (14.57) | 2 | 63 | 39.38 | 4 | 21.80 | 15 |
| Liguria | 1 | 59 (2.98) | 10 | 15 | 23 (1.16) | 8 | 7 | 11.86 | 13 | 30.43 | 13 |
| Lombardy | 9 | 403 (20.32) | 1 | 166 | 769 (38.78) | 1 | 233 | 57.82 | 1 | 30.30 | 14 |
| Marche | 3 | 37 (1.87) | 14 | 6 | 8 (0.40) | 14 | 6 | 16.22 | 10 | 75.00 | 1 |
| Molise | 0 | 0 (0.00) | 19 | 2 | 2 (0.10) | 17 | NA | NA | NA | NA | NA |
| Piedmont | 3 | 134 (6.76) | 6 | 51 | 147 (7.41) | 6 | 57 | 42.54 | 3 | 38.78 | 9 |
| Sardinia | 2 | 25 (1.26) | 15 | 5 | 9 (0.45) | 13 | 3 | 12.00 | 12 | 33.33 | 10 |
| Sicily | 3 | 118 (5.95) | 7 | 12 | 85 (4.29) | 7 | 62 | 52.54 | 2 | 72.94 | 4 |
| Tuscany | 4 | 213 (10.74) | 3 | 48 | 215 (10.84) | 3 | 67 | 31.46 | 5 | 31.16 | 12 |
| Trentino AA | 1 | 20 (1.01) | 16 | 5 | 10 (0.50) | 12 | 2 | 10.00 | 14 | 20.00 | 16 |
| Umbria | 1 | 59 (2.98) | 10 | 4 | 4 (0.20) | 15 | 3 | 5.08 | 16 | 75.00 | 1 |
| Veneto | 3 | 170 (8.57) | 4 | 36 | 155 (7.82) | 5 | 50 | 29.41 | 7 | 32.26 | 11 |
| Total/mean | 58 | 1,983 (100) | - | 509 | 1,983 (100) | - | 690 | 42.35[iii] | - | 39.34[iii] | - |

*Table 1: Market for university-industry research collaboration, by region*

[i] Number of universities that realized at least one university-enterprise collaboration (supply)

[ii] Number of enterprises that realized at least one university-enterprise collaboration (demand)

[iii] The weighted mean of the regional data, with weighting equal to the share of regional university-industry collaborations with respect to the national total

# 4. University performance in collaboration with industry: analysis by scientific sector

The preceding regional level analysis of university-enterprise collaborations does not distinguish the scientific sectors involved in the collaborations. As previously noted, the 1,534 publications in the dataset actually represent 2,363 collaborations at the SDS-enterprise level, in a total of 141 SDS. For each SDS, it is possible to characterize the supply of knowledge by each university, on the basis of the number of collaborations achieved with enterprises (SDS-company collaborations), both at the intra-regional and extra-regional levels (internal and external markets). In the same fashion, the sectorial demand for collaboration can be characterized in terms of SDS-company collaborations, by region.

## 4.1 *Degree of correspondence between regional demand and supply*

The capacity of universities to transfer new knowledge to enterprises depends in part on the intensity of demand they encounter. Characterizing the demand and supply of knowledge by region, in each scientific sector, could be useful for all three types of actors: enterprises, universities and policy makers. Such information would assist enterprises to make strategic choices concerning the location of their activities. It could also be useful for universities to understand regional demand, in order to adjust their capacity to satisfy such demand. Information on the balance between supply and demand in different sectors is also useful for the regional policy maker, enabling the selection of interventions that would bring balance and increase regional benefits from knowledge transfer. Naturally, the interests of the various actors may differ. Universities, in their search for external funding, will not ignore the potential of collaboration with enterprises in other regions, even though they will be at some competitive disadvantage compared to closer universities, due to the proximity effect. Other factors being equal, enterprises will prefer to collaborate with local universities, in order to minimize costs, and possibly in order to exploit incentives offered by their home region. Meanwhile, regional policy makers will have an interest that the positive externalities of knowledge transfer fall in the region, which may imply an interest in limiting the exit of new knowledge produced in their region, to the benefit of "competing" private enterprises situated elsewhere. The comparison of industry demand versus the actual supply of collaboration from universities, for each SDS, permits us to measure the sectorial levels of university-enterprise correspondence, by region. Inversely, it is also possible to measure the degree of correspondence between the total supply by the universities of a particular region relative to the demand from local enterprises.

Analysis of the dataset shows that, on average, 1.99 university researchers are involved in each university-industry co-publication. This means that universities, to satisfy a demand for collaboration received from an enterprise, will need an average of at least two researchers. Thus, in formal terms, the degree of correspondence between regional demand and the supply from a generic university $i$, in a specific SDS, can be so defined:

(university degree of correspondence)$_i$ = $\alpha$*num_researchers$_i$ - annual regional demand$_m$

where:
- α is a coefficient taking into account how many academic researchers, on average, are involved in each university-industry publication[12]
- $num\_researchers_i$ = number of researchers at university $i$[13]
- annual regional $demand_m$ = mean annual number of collaborations between enterprises in region $m$ and universities throughout Italy.

Similarly, the degree of correspondence can be also defined for a generic region $m$:

(region degree of correspondence)$_m$ = α*$num\_researchers_m$ - annual regional $demand_m$

where:
- $num\_researchers_m$ = number of researchers affiliated to universities of region $m$

Using these two indicators, values greater than zero indicate a surplus (excess of supply), while values less than zero indicate a shortage.

Comparing between regions, higher values of shortage indicate a higher level of industrial demand with respect to university supply and thus, other factors being equal, a higher probability that internal demand will be satisfied by supply from extra-regional universities. Higher values of surplus represent an excess of supply with respect to demand, meaning that it will be difficult for the production of new knowledge to result in economic benefits only within the originating region, and that benefits will likely also accrue in other regions.

As an example of an analysis, we have chosen the biochemistry SDS, which achieves a notable number of university-industry co-publications (89) in the period under observation. Table 2 shows the degree of correspondence between industry demand and university supply for the biochemistry SDS in Italy. The table shows the situation for the 49 universities that represent the supply in this sector, meaning that they have at least one researcher that belongs to the biochemistry SDS. For this SDS, the dataset contains 89 articles published in university-enterprise co-authorship, representing a total of 98 SDS-enterprise collaborations, or an average of 33 per year. In Molise (and as previously seen for all sectors in Valle d'Aosta) both demand and supply are nil, and therefore this region is not included in Table 2. The remaining 18 regions all present a supply from universities, but only nine regions register a demand for collaboration: Abruzzo, Emilia Romagna, Friuli Veneto Giulia, Lazio, Liguria, Lombardy, Piedmont, Tuscany and Veneto. There is no region where enterprise demand encounters a null offer of supply.

On the supply side, of the 49 universities with researchers in biochemistry, only 27 actually achieved collaborations with enterprises. The highest level of supply is from University of Rome "Sapienza" (7.41% of the national total), University of Milan (6.94%), University of Bologna (5.79%) and University of Naples "Federico II" (5.79%). Twenty of the 49 universities with potential for collaboration are situated in regions with a null demand. These universities thus cannot contribute to local regional development through collaboration with private enterprises involved in this particular sector[14]. Among these universities, those with the highest number of researchers, and thus the highest value of surplus supply (regional demand being null), are the University of Naples "Federico II" (50 researchers), University of Bari (38) and the Second University of Naples (24).

---

[12] The previous finding that, on average, each university-industry publication is authored by two academics allows to consider α equal to 1/2. This number can readily be adjusted according to the characteristics of each specific SDS, to appropriately adapt the analysis.

[13] The figures for numbers of scientists in each SDS, averaged over the triennium under consideration, were obtained from a database of the Ministry of Education, Universities and Research.

---

[14] This interpretation is not intended as a superficial suggestion that universities should resize their research capacity in the SDS examined. Capacity must also be planned in relation to the other two primary roles of the university: higher education and research.

| University ID [i] | University name | Region | Number of university scientists (% of total) | Annual regional demand[ii] (% of total) | Degree of correspondence | |
|---|---|---|---|---|---|---|
| | | | | | University level[iii] | Regional level[iv] |
| 1 | Polytechnic University of Ancona | Marche | 17 (1.97) | 0.00 (0.00) | 8.50 | 23.50 |
| 2 | Sacred Heart Catholic University | Lombardy | 16 (1.85) | 8.00 (24.49) | 0.00 | 58.50 |
| 3 | Scuola Superiore Sant'Anna of Pisa | Tuscany | 1 (0.12) | 7.67 (23.47) | -7.17 | 28.33 |
| 4 | Second University of Naples | Campania | 24 (2.78) | 0.00 (0.00) | 12.00 | 41.50 |
| 5 | Roma University Tre | Lazio | 3 (0.35) | 4.33 (13.27) | -2.83 | 47.17 |
| 6 | University of Bari | Apulia | 38 (4.40) | 0.00 (0.00) | 19.00 | 22.50 |
| 7 | University of Basilicata | Basilicata | 3 (0.35) | 0.00 (0.00) | 1.50 | 1.50 |
| 8 | University of Benevento 'Sannio' | Campania | 2 (0.23) | 0.00 (0.00) | 1.00 | 41.50 |
| 9 | University of Bologna | Emilia Romagna | 50 (5.79) | 2.67 (8.16) | 22.33 | 48.83 |
| 10 | University of Brescia | Lombardy | 13 (1.50) | 8.00 (24.49) | -1.50 | 58.50 |
| 11 | University of Cagliari | Sardinia | 13 (1.50) | 0.00 (0.00) | 6.50 | 12.00 |
| 12 | University of Calabria | Calabria | 5 (0.58) | 0.00 (0.00) | 2.50 | 8.00 |
| 13 | University of Camerino | Marche | 11 (1.27) | 0.00 (0.00) | 5.50 | 23.50 |
| 14 | University of Catania | Sicily | 20 (2.31) | 0.00 (0.00) | 10.00 | 26.50 |
| 15 | University of Catanzaro 'Magna Grecia' | Calabria | 11 (1.27) | 0.00 (0.00) | 5.50 | 8.00 |
| 16 | University of Chieti 'Gabriele D'Annunzio' | Abruzzo | 11 (1.27) | 0.33 (1.02) | 5.17 | 13.67 |
| 17 | University of Eastern Piedmont 'A. Avogadro' | Piedmont | 5 (0.58) | 1.33 (4.08) | 1.17 | 14.67 |
| 18 | University of Ferrara | Emilia Romagna | 17 (1.97) | 2.67 (8.16) | 5.83 | 48.83 |
| 19 | University of Florence | Tuscany | 23 (2.66) | 7.67 (23.47) | 3.83 | 28.33 |
| 20 | University of Foggia | Apulia | 2 (0.23) | 0.00 (0.00) | 1.00 | 22.50 |
| 21 | University of Genova | Liguria | 17 (1.97) | 0.33 (1.02) | 8.17 | 8.17 |
| 22 | University of L'Aquila | Abruzzo | 13 (1.50) | 0.33 (1.02) | 6.17 | 13.67 |
| 23 | University of Lecce | Apulia | 5 (0.58) | 0.00 (0.00) | 2.50 | 22.50 |
| 24 | University of Messina | Sicily | 14 (1.62) | 0.00 (0.00) | 7.00 | 26.50 |
| 25 | University of Milan | Lombardy | 60 (6.94) | 8.00 (24.49) | 22.00 | 58.50 |
| 26 | University of Milan 'Bicocca' | Lombardy | 12 (1.39) | 8.00 (24.49) | -2.00 | 58.50 |
| 27 | University of Modena and Reggio Emilia | Emilia Romagna | 15 (1.74) | 2.67 (8.16) | 4.83 | 48.83 |
| 28 | University of Naples 'Federico II' | Campania | 50 (5.79) | 0.00 (0.00) | 25.00 | 41.50 |
| 29 | University of Padua | Veneto | 36 (4.17) | 5.67 (17.35) | 12.33 | 19.83 |
| 30 | University of Palermo | Sicily | 19 (2.20) | 0.00 (0.00) | 9.50 | 26.50 |
| 31 | University of Parma | Emilia Romagna | 21 (2.43) | 2.67 (8.16) | 7.83 | 48.83 |
| 32 | University of Pavia | Lombardy | 23 (2.66) | 8.00 (24.49) | 3.50 | 58.50 |

| University ID [i] | University name | Region | Number of university scientists (% of total) | Annual regional demand [ii] (% of total) | Degree of correspondence University level [iii] | Degree of correspondence Regional level [iv] |
|---|---|---|---|---|---|---|
| 33 | University of Perugia | Umbria | 23 (2.66) | 0.00 (0.00) | 11.50 | 11.50 |
| 34 | University of Pisa | Tuscany | 29 (3.36) | 7.67 (23.47) | 6.83 | 28.33 |
| 35 | University of Rome 'La Sapienza' | Lazio | 64 (7.41) | 4.33 (13.27) | 27.67 | 47.17 |
| 36 | University of Rome 'Tor Vergata' | Lazio | 29 (3.36) | 4.33 (13.27) | 10.17 | 47.17 |
| 37 | University of Salerno | Campania | 7 (0.81) | 0.00 (0.00) | 3.50 | 41.50 |
| 38 | University of Sassari | Sardinia | 11 (1.27) | 0.00 (0.00) | 5.50 | 12.00 |
| 39 | University of Siena | Tuscany | 19 (2.20) | 7.67 (23.47) | 1.83 | 28.33 |
| 40 | University of Teramo | Abruzzo | 4 (0.46) | 0.33 (1.02) | 1.67 | 13.67 |
| 41 | University of Trento | Trentino Alto Adige | 1 (0.12) | 0.00 (0.00) | 0.50 | 0.50 |
| 42 | University of Trieste | Friuli V. Giulia | 18 (2.08) | 2.33 (7.14) | 6.67 | 12.67 |
| 43 | University of Turin | Piedmont | 27 (3.13) | 1.33 (4.08) | 12.17 | 14.67 |
| 44 | University of Udine | Friuli V. Giulia | 12 (1.39) | 2.33 (7.14) | 3.67 | 12.67 |
| 45 | University of Urbino 'Carlo Bo' | Marche | 19 (2.20) | 0.00 (0.00) | 9.50 | 23.50 |
| 46 | University of Varese 'Insubria' | Lombardy | 9 (1.04) | 8.00 (24.49) | -3.50 | 58.50 |
| 47 | University of Venice 'Ca' Foscari' | Veneto | 3 (0.35) | 5.67 (17.35) | -4.17 | 19.83 |
| 48 | University of Verona | Veneto | 12 (1.39) | 5.67 (17.35) | 0.33 | 19.83 |
| 49 | University of Viterbo 'Tuscia' | Lazio | 7 (0.81) | 4.33 (13.27) | -0.83 | 47.17 |
| Mean | - | - | 17.63 | 2.70 | 6.12 | 29.56 |

*Table 2: Degree of correspondence between industry demand and university supply of research collaboration for the biochemistry SDS*

[i] Each university is given an identification number to simplify reference throughout subsequent analysis.

[ii] Annual regional demand is the average number of collaborations per year by enterprises in the same region (the region of the university examined) with universities throughout Italy.

[iii] The surplus at the level of an individual university is calculated as the difference between half the number of scientists of the university and the annual regional demand.

[iv] Regional surplus is calculated as the difference between half the number of scientists of all the universities in a region and the annual regional demand.

The highest levels of regional demand are seen in Lombardy (24.49% of the national total), Tuscany (23.47%) and Veneto (17.35%). Examining the degree of correspondence between single university supply and regional demand, the mean value results as a surplus of 6.12. The greatest shortage (-7.17) is recorded for the Scuola Superiore Sant'Anna of Pisa: given a high level of regional demand for collaboration (Tuscany, 7.67) this institution has only one researcher in the SDS examined. The highest surplus of supply (+27.67) is seen for the University of Rome "Sapienza", where there is a low annual regional demand (4.33) but there are a full 64 university researchers in biochemistry.

The overall degrees of correspondence at the regional level, given by the difference between half the total number of researchers in all the universities and the annual regional demand, shows that all the regions have a surplus of supply relative to local demand. The highest levels of surplus are seen in Lombardy (58.50), Emilia Romagna (48.83) and Lazio (47.17), while the lowest surpluses are seen in Calabria (8.00), Basilicata (1.50) and Trento Alto Adige (0.50).

University managers can use this information to appropriately adjust the supply of research from their institutions, while also considering the needs for its education program. Management would need to consider the degree of correspondence between its own university's supply and regional demand, the general intensity of local competition (surplus at the regional level) and the capacity of the competition to capture local demand (see next section of this paper). Regional policy-makers can also use results from this type of analysis in a similar way, to formulate polices and measures that develop better correspondence between demand and supply, in keeping with overall regional strategies for development.

### 4.2 *Measurement of collaboration performance by individual universities*

This section presents an example of how the model can be applied to the comparative measurement of universities' performance in terms of their capacity to participate in research collaborations with industry. This aspect of the model requires several assumptions that may eventually be abandoned, in full or in part, based on further development of the methodology.

A university's capacity to capture enterprise demand for collaboration depends on various factors. Some of these factors concern the actual merit of the university, among these being its scientific excellence and its related prestige, and its effort and tendency to collaborate with the private sector. However, some factors do not depend on merit, instead involving conditions such as geographic and social proximity[15], and potential regional incentives in favour of organizations and collaborations on its territory. Given equal merit and social proximity, the effect of geographic proximity should induce enterprises in a given region to prefer collaborations with local universities. Under these conditions, it would be expected that the advantage of location will permit universities of a given region to capture the entire demand of local enterprises, each university in proportion to their share of the regional supply. To calculate this share we can assume that each research scientist is able to satisfy one demand for collaboration per year, in the SDS to which the scientist belongs, independently of the specific content of the collaboration. The model also assumes that: i) differences in distance from an enterprise to the universities in a single region bear no influence on the enterprise's choice of local research partner; ii) extra-regional collaborations are not influenced by distance between the partners involved. This final assumption is equivalent to considering that if an enterprise selects not to collaborate with a local university then it views all extra-regional universities as being equal in the sense of geographic distance[16].

---
[15] The term "social proximity" is used in the sense given by Boschma (2005): "Social proximity is defined here in terms of embedded relations between agents at the micro-level. Relations between actors are socially embedded when they involve trust based on friendship, kinship and experience".

The first step is to develop a measure of the capacity of universities to capture local demand for collaboration, followed by a measure of their capacity to capture the extra-regional demand. The analysis must be conducted at the level of each SDS, since each sector is expected to have different characteristics of demand, supply and intervention by local governments.

For each SDS, the model compares the mean annual number of research collaborations carried out by a university (annual intra-regional supply) to the number expected for the university (expected annual intra-regional supply), given the demand for collaboration by local enterprises and the supply from "competing" universities in the same region. The expected annual regional supply is given by the annual regional demand multiplied by the university's potential share of the total regional supply (number of scientists at the university in the specific SDS divided by total number of university scientists in the region in the same SDS). The ratio of number of collaborations achieved and number of collaborations expected provides the measure of the university's capacity to capture internal demand for collaboration, which we term "intra-regional performance".

In formal terms, for a given SDS and the generic university $i$:

(expected annual intra-regional supply)$_i =$

$$= \text{annual regional demand}_{m_i} \cdot \frac{\text{num\_researchers}_i}{\text{num\_regional\_researchers}_{m_i}}$$

where:
- $m_i =$ is the region where the university $i$ is situated
- annual regional demand$_m$ = mean annual number of collaborations between enterprises in region $m$ and universities throughout Italy
- num\_researchers$_i$ = number of researchers at university $i$
- num\_regional\_researchers$_{m_i}$ = total of researchers at all the universities in region $m$.

Hence:
Intra-regional performance =

=Annual intra-regional supply / Expected annual intra-regional supply

A university will have a null value for the indicator of annual intra-regional supply when annual regional demand is null or when its number of researchers is null. Thus, in essence, if the university does not have researchers in the SDS examined or is situated in a region with no demand for collaboration then it is not meaningful to further

---
[16] Further development of the proposed model could remove this assumption and take dues consideration of the effect of different distances.

measure intra-regional performance.

Next, the capacity of a university to capture extra-regional demand for collaborations can be represented by comparing the mean annual number of its collaborations realized with extra-regional enterprises (annual extra-regional supply) and the number of collaborations expected (expected annual extra-regional supply). The value of annual extra-regional supply is obtained by considering, for all regions, the annual extra-regional demand for collaboration that each university, given its share of the total national researchers, could be expected to capture.

Formally, for a given SDS and the generic university $i$:

(Expected annual extra-regional supply)$_i$ =

$$= \sum_{m=1, m \neq m_i}^{M} \left[ \text{Annual extra-regional demand}_m \times \frac{\text{num\_researchers}_i}{\sum_{t=1, t \neq m}^{M} \text{num\_regional\_researchers}_t} \right]$$

where:
- $m_i$ = is the region where the university is situated
- $M$ = the number of regions in Italy (20)
- annual extra-regional demand$_m$ = mean annual number of collaborations by enterprises from region $m$ with extra-regional universities
- num\_researchers$_i$ = number of researchers at university $i$
- num\_regional\_researchers$_t$ = number of researchers at the universities in region $t$.

Hence:
Extra-regional performance =
= Annual extra-regional supply / Expected annual extra-regional supply

Finally, as a step towards a single indicator of overall performance that measures overall capacity of a university to realize collaborations with the industrial sector, the model alsocalculates the mean value of intra-regional and extra-regional performance[17].

Returning to the same sector previously used as an illustration, Table 3 presents measures of performance for the 49 universities with at least one researcher in the biochemistry SDS. Only 27 of these universities actually collaborated with enterprises: two collaborated only with local enterprises; 14 only with extra-regional enterprises; the remaining eleven with both intra and extra-regional enterprises.

---

[17] The greater difficulty involved in activating extra-regional compared to intra-regional collaborations could be considered by assigning a higher weight to extra-regional performance in the calculation of overall performance.

| University ID | University | Annual regional demand | Annual intra-regional supply[i] | Expected annual intra-regional supply[ii] | Intra-regional performance[iii] | Annual extra-regional supply[iv] | Expected annual extra-regional supply[v] | Extra-regional performance[vi] | Overall performance[vii] |
|---|---|---|---|---|---|---|---|---|---|
| 1 | Polytechnic University of Ancona | 0.00 | NA | 0.00 | NA | 0.33 | 0.39 | 0.85 | NA |
| 2 | Sacred Heart Catholic University | 8.00 | 0.00 | 0.96 | 0.00 | 0.00 | 0.23 | 0.00 | 0.00 |
| 3 | Scuola Superiore Sant'Anna of Pisa | 7.67 | 0.00 | 0.11 | 0.00 | 0.00 | 0.02 | 0.00 | 0.00 |
| 4 | Second University of Naples | 0.00 | NA | 0.00 | NA | 1.33 | 0.55 | 2.42 | NA |
| 5 | University of 'Roma Tre' | 4.33 | 0.00 | 0.13 | 0.00 | 0.00 | 0.06 | 0.00 | 0.00 |
| 6 | University of Bari | 0.00 | NA | 0.00 | NA | 0.00 | 0.87 | 0.00 | NA |
| 7 | University of Basilicata | 0.00 | NA | 0.00 | NA | 0.33 | 0.07 | 4.83 | NA |
| 8 | University of Benevento 'Sannio' | 0.00 | NA | 0.00 | NA | 0.00 | 0.05 | 0.00 | NA |
| 9 | University of Bologna | 2.67 | 0.33 | 1.29 | 0.26 | 0.67 | 1.11 | 0.60 | 0.43 |
| 10 | University of Brescia | 8.00 | 0.00 | 0.78 | 0.00 | 0.00 | 0.19 | 0.00 | 0.00 |
| 11 | University of Cagliari | 0.00 | NA | 0.00 | NA | 0.00 | 0.30 | 0.00 | NA |
| 12 | University of Calabria | 0.00 | NA | 0.00 | NA | 0.00 | 0.11 | 0.00 | NA |
| 13 | University of Camerino | 0.00 | NA | 0.00 | NA | 0.33 | 0.25 | 1.32 | NA |
| 14 | University of Catania | 0.00 | NA | 0.00 | NA | 0.33 | 0.46 | 0.73 | NA |
| 15 | University of Catanzaro 'Magna Grecia' | 0.00 | NA | 0.00 | NA | 0.00 | 0.25 | 0.00 | NA |
| 16 | University of Chieti 'Gabriele D'Annunzio' | 0.33 | 0.00 | 0.13 | 0.00 | 0.00 | 0.25 | 0.00 | 0.00 |
| 17 | University of Eastern Piedmont 'A. Avogadro' | 1.33 | 0.00 | 0.21 | 0.00 | 0.33 | 0.11 | 3.12 | 1.56 |
| 18 | University of Ferrara | 2.67 | 1.00 | 0.44 | 2.27 | 0.33 | 0.38 | 0.89 | 1.58 |
| 19 | University of Florence | 7.67 | 0.33 | 2.45 | 0.14 | 0.33 | 0.39 | 0.85 | 0.49 |
| 20 | University of Foggia | 0.00 | NA | 0.00 | NA | 0.00 | 0.05 | 0.00 | NA |
| 21 | University of Genova | 0.33 | 0.00 | 0.33 | 0.00 | 1.67 | 0.38 | 4.34 | 2.17 |
| 22 | University of L'Aquila | 0.33 | 0.00 | 0.15 | 0.00 | 1.33 | 0.29 | 4.54 | 2.27 |
| 23 | University of Lecce | 0.00 | NA | 0.00 | NA | 0.00 | 0.11 | 0.00 | NA |
| 24 | University of Messina | 0.00 | NA | 0.00 | NA | 0.33 | 0.32 | 1.04 | NA |
| 25 | University of Milan | 8.00 | 1.00 | 3.61 | 0.28 | 1.00 | 0.86 | 1.16 | 0.72 |
| 26 | University of Milan 'Bicocca' | 8.00 | 0.33 | 0.72 | 0.46 | 0.00 | 0.17 | 0.00 | 0.23 |
| 27 | University of Modena and Reggio Emilia | 2.67 | 0.67 | 0.39 | 1.72 | 0.67 | 0.33 | 2.01 | 1.86 |
| 28 | University of Naples 'Federico II' | 0.00 | NA | 0.00 | NA | 1.33 | 1.15 | 1.16 | NA |

| University ID | University | Annual regional demand | Annual intra-regional supply[i] | Expected annual intra-regional supply[ii] | Intra-regional performance[iii] | Annual extra-regional supply[iv] | Expected annual extra-regional supply[v] | Extra-regional performance[vi] | Overall performance[vii] |
|---|---|---|---|---|---|---|---|---|---|
| 29 | University of Padua | 5.67 | 0.00 | 4.00 | 0.00 | 0.00 | 0.80 | 0.00 | 0.00 |
| 30 | University of Palermo | 0.00 | NA | 0.00 | NA | 0.00 | 0.44 | 0.00 | NA |
| 31 | University of Parma | 2.67 | 0.00 | 0.54 | 0.00 | 0.33 | 0.46 | 0.72 | 0.36 |
| 32 | University of Pavia | 8.00 | 0.33 | 1.38 | 0.24 | 0.00 | 0.33 | 0.00 | 0.12 |
| 33 | University of Perugia | 0.00 | NA | 0.00 | NA | 0.33 | 0.53 | 0.63 | NA |
| 34 | University of Pisa | 7.67 | 0.33 | 3.09 | 0.11 | 0.33 | 0.50 | 0.67 | 0.39 |
| 35 | University of Rome 'La Sapienza' | 4.33 | 1.00 | 2.69 | 0.37 | 1.00 | 1.22 | 0.82 | 0.60 |
| 36 | University of Rome 'Tor Vergata' | 4.33 | 0.33 | 1.22 | 0.27 | 1.00 | 0.55 | 1.81 | 1.04 |
| 37 | University of Salerno | 0.00 | NA | 0.00 | NA | 0.00 | 0.16 | 0.00 | NA |
| 38 | University of Sassari | 0.00 | NA | 0.00 | NA | 0.00 | 0.25 | 0.00 | NA |
| 39 | University of Siena | 7.67 | 2.33 | 2.02 | 1.15 | 1.67 | 0.32 | 5.13 | 3.14 |
| 40 | University of Teramo | 0.33 | 0.00 | 0.05 | 0.00 | 0.00 | 0.09 | 0.00 | 0.00 |
| 41 | University of Trento | 0.00 | NA | 0.00 | NA | 0.00 | 0.02 | 0.00 | NA |
| 42 | University of Trieste | 2.33 | 2.00 | 1.40 | 1.43 | 0.33 | 0.41 | 0.82 | 1.12 |
| 43 | University of Turin | 1.33 | 0.00 | 1.13 | 0.00 | 0.00 | 0.58 | 0.00 | 0.00 |
| 44 | University of Udine | 2.33 | 0.00 | 0.93 | 0.00 | 1.33 | 0.27 | 4.92 | 2.46 |
| 45 | University of Urbino 'Carlo Bo' | 0.00 | NA | 0.00 | NA | 0.33 | 0.44 | 0.76 | NA |
| 46 | University of Varese 'Insubria' | 8.00 | 0.00 | 0.54 | 0.00 | 0.00 | 0.13 | 0.00 | 0.00 |
| 46 | University of Venice 'Ca' Foscari' | 5.67 | 0.00 | 0.33 | 0.00 | 0.00 | 0.07 | 0.00 | 0.00 |
| 48 | University of Verona | 5.67 | 5.00 | 1.33 | 3.75 | 0.33 | 0.27 | 1.25 | 2.50 |
| 49 | University of Viterbo 'Tuscia' | 4.33 | 0.00 | 0.29 | 0.00 | 0.00 | 0.13 | 0.00 | 0.00 |

*Table 3: Performance by universities in R&D collaborations for the biochemistry SDS*

i Annual intra-regional supply is defined as the mean annual number of collaborations by a university with the enterprises in its home region.

ii Expected annual intra-regional supply is defined as annual regional demand multiplied by the university's share of the regional total of researchers for the SDS examined.

iii Intra-regional performance = annual intra-regional supply/expected annual intra-regional supply.

iv Annual extra-regional supply is the mean annual number of collaborations by the university with extra-regional enterprises.

v Expected annual extra-regional supply for a given university is given by the sum, for all regions, of mean annual demand from extra-regional enterprise that is not captured by local universities, multiplied by the number of researchers at the university relative to the total of extra-regional academic researchers. See the text for the relevant formula.

vi Extra-regional performance = annual extra-regional supply / expected annual extra-regional supply.

vii Overall performance = (intra-regional performance + extra-regional performance)/2.

From the side of intra-regional supply, of the 49 universities that show research capacity, 20 encounter a null local demand. However, of these 20 universities, 9 succeeded in capturing extra-regional demand. The remaining 29 universities are situated in regions with a local demand for collaboration, but of these, eleven did not collaborate with industry. Of the 18 universities that were involved in collaborations, five did so only with extra-regional enterprises. Thus an overall total of 16 (11+5) universities did not exploit their advantage of location.

The universities with the best intra-regional performance in collaboration are the University of Verona (intra-regional performance of 3.75), University of Ferrara (2.27) and the University of Modena and Reggio Emilia (1.72). The universities with the highest extra-regional performance are the University of Siena (5.13), University of Udine (4.92) and University of Basilicata (4.83). For overall performance, the highest ranking institutions are the University of Siena (3.14), University of Verona (2.50) and University of Udine (2.46).

Figure 1 permits a clearer comparison of intra and extra-regional performance. Given the preceding definitions, the universities that will present values for both types of performance are the 29 situated in the nine regions with local demand. Of these 29, the universities with the poorest performance are those eleven that, even though facing a regional demand, did not collaborate at either the intra or extra-regional level (null values of intra and extra-regional performance). The figure positions the remaining 18 universities, which are both: i) located in a region where there is demand and ii) achieved at least one collaboration. Five universities are located along the vertical axis (extra-regional performance), since although they faced a regional demand, they succeeded in collaborating only with extra-regional enterprises. Two universities are located along the horizontal axis (intra-regional performance), since they did not achieve any collaborations outside their region. Eleven universities show positive values for both dimensions of performance.

Graphing the medians for the distributions (0.27 for the x axis and 1.03 for the y axis), quadrant II then contains the universities that show the highest performance in collaborations with industry at both the intra and extra-regional levels. Quadrant III again contains universities from the group of 18 with collaborations at intra and extra-regional levels, but here at a lower level of performance. The best performing universities, in quadrant II, are the University of Verona (University ID 48), University of Modena and Reggio Emilia (ID 27) and University of Siena (ID 39), along with the University of Rome "Tor Vergata" (ID 36), which has a value for intra-regional performance at the national median level. The University of Milan (ID 25) is positioned in quadrant II, but almost at the intersection of the two axes. Most of these specifically named universities are located in centre-north Italy, especially in industrially advanced regions. They are universities that do better than others in satisfying not only local demand, but also that from other regions.

As noted, quadrant III contains universities with both intra and extra-regional collaborations, but at lower levels of performance. These are the University of Parma (ID 31), University of Pisa (ID 34), University of Pavia (ID 32), University of Bologna (ID 9) and University of Florence (ID 19). It is interesting to observe that four of these five universities are situated in two regions (Emilia Romagna and Tuscany) that also feature two of the top performing universities of quadrant II.

Quadrant I contains universities that, although they achieve good performance at the extra-regional level, do not achieve similar levels of performance at the local level.

These are the University of Udine (ID 44), University of Eastern Piedmont "Amedeo Avogadro" (ID 17), University of Aquila (ID 22) and University of Genova (ID 21). These universities are situated in regions characterized by low levels of annual regional demand: Friuli Venezia Giulia (2.33), Piedmont (1.33), Abruzzo (0.33) and Liguria (0.33). Quadrant IV contains universities that demonstrated good performance at the local level but low performance at the extra-regional level: University of Ferrara (ID 18), University of Trieste (ID 42), University of Rome "Sapienza" (ID 35) and University of Milan "Bicocca" (ID 26). Their home regions (Emilia Romagna, Tuscany, Lazio, Lombardy) are clearly among the most industrially developed in Italy, with high values of regional demand (respectively 2.67, 7.67, 4.33 and 8.00).

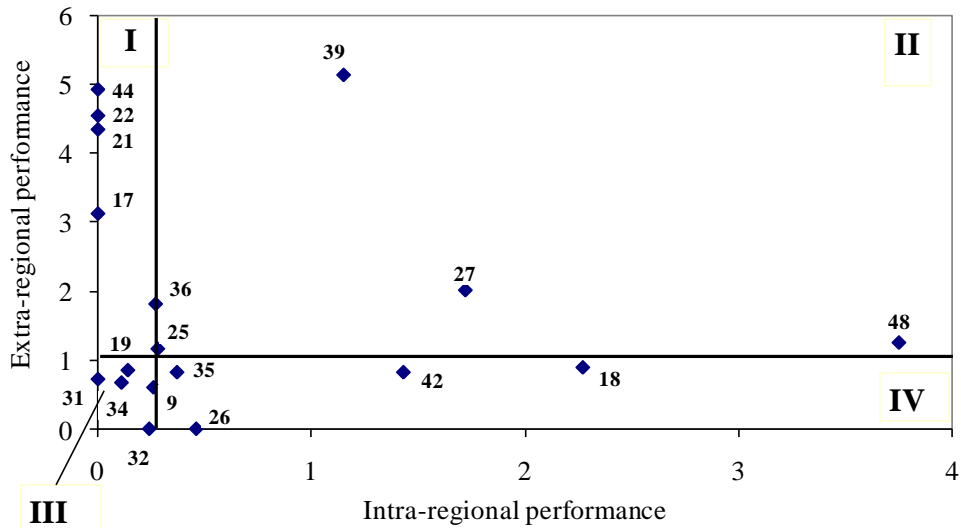

*Figure 1: Positioning of Italian universities in terms of intra-regional and extra-regional performance for collaborations in the biochemistry SDS. Universities with no collaborations of either type are not shown.*

The 20 universities situated in regions with no demand are not included in the figure because the values of intra-regional performance are undefined, but of these 20, nine did succeed in capturing a part of the extra-regional demand.

The exercise can be repeated for all SDS. In addition to other potentially useful information, management of a university can draw on this particular aspect to compare the positioning of their own institution with respect to those of the other universities in their region and with respect to all others in Italy. Management could also conduct time series analysis to measure the effectiveness of policies and initiatives intended to promote collaboration both within and beyond the home region. In the same ways, national policy makers can draw on the information to evaluate performance in university-industry collaboration, but caution must be exercised in interpreting data relative to universities in different regions: higher intra-regional performance may originate not from the merits of the individual universities, but actually be more due to effective policies and initiatives enacted by regional governments.

### 4.3 *The regional perspective*

The above data can also be analyzed from a regional perspective, useful to the local policy maker. As a potential illustration, Figure 2 shows the results for the region of Lazio in the biochemistry SDS. In the period under examination, seven enterprises situated in Lazio collaborated with universities, realizing 13 SDS-enterprise collaborations: four were with local universities and the remaining nine with universities from other regions (Campania, Sicily, Abruzzo, Lombardy and Tuscany). There are four Lazio-based universities active in this SDS, but of these only two collaborated with industry (University of Rome "Tor Vergata", with 29 research scientists; University of Rome "Sapienza", 64 scientists). In further detail, the University of Rome "Tor Vergata" collaborated with one local enterprise and three extra-regional enterprises (situated in Tuscany and Friuli Venezia Giulia), while University of Rome "Sapienza" collaborated with three local enterprises and three extra-regional enterprises (in Piedmont and Tuscany).

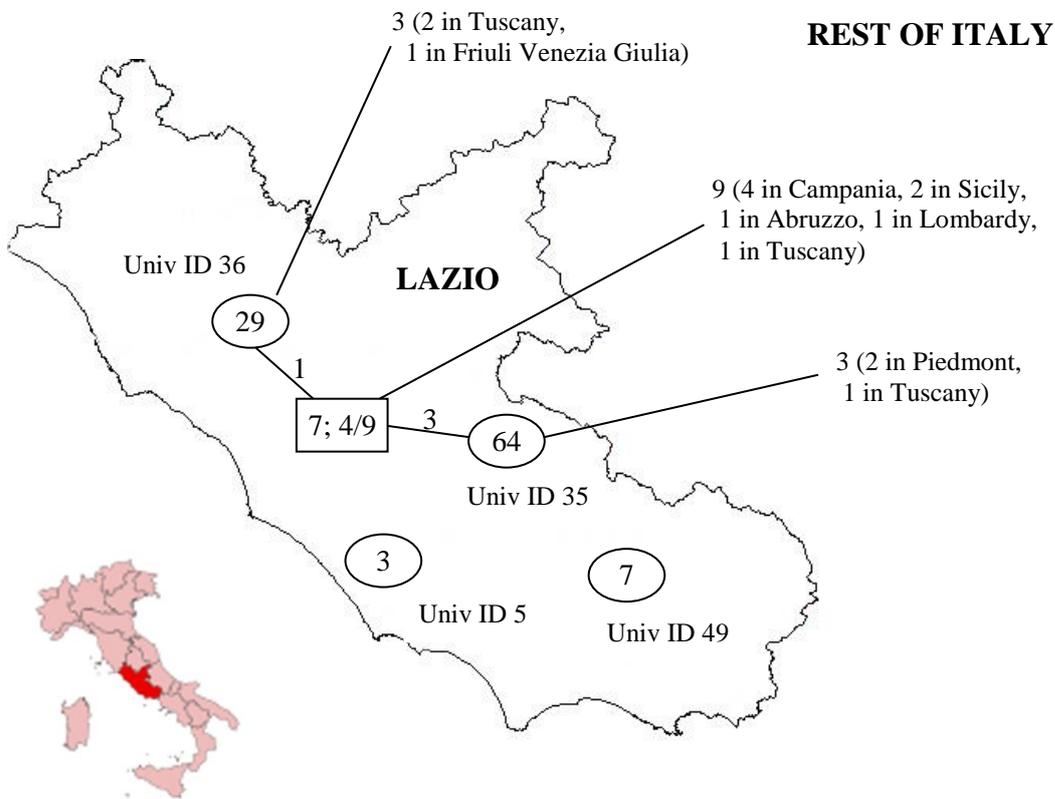

*Figure 2: Collaborations involving universities and enterprises in Lazio, for the biochemistry SDS.*

Legend

| $f; n/m$ | Enterprises: $f$ is the number of enterprises that collaborated with universities; $n$ is the number of SDS-enterprise intra-regional collaborations; $m$ is the number of SDS-enterprise extra-regional collaborations. |

( $s$ ) University active in the SDS examined, with s indicating the number of scientists that belong to the SDS

—c— c is the number of collaborations between the two connected entities

## 5. Conclusions

PRIs are increasingly called on to better implement their role in supporting regional and national socio-economic development. This necessitates the parallel development of measures and actual instruments to measure their performance. National PRI technology transfer assessments have essentially been performed in the dimension of licensing and spin-offs, because of availability of data. But the aspect of research collaboration as a means of technology transfer is of great interest, and would at least be equally significant to measure. This work proposes a bibliometric measurement framework for comparing the performance of universities in participating in research collaborations with industry, which includes the feature of accounting for possible advantages due to location. The conduct of the analysis at the level of individual scientific disciplines and regions avoids the inevitable distortions that would result from regional aggregation, and also disciplinary aggregation. This detailed level of analysis also permits subsequent development of policy and interventions for specific disciplines and/or regions.

The results from applying the model should be of interest to: i) university management, for strategic and operational control in the dimension of research collaboration with industry; ii) regional decision-makers, for formulation of policies and stimuli initiatives towards greater university-industry integration; iii) national decision-makers, in conducting evaluations of technology transfer and in coordination and integration of regional innovation systems.

Future development of the model proposed could concern the inclusion of joint university-enterprise patents as a further proxy of public-private research collaboration, particularly in nations where the numbers of patents offer a significant complement to data concerning joint scientific publications. In addition, the actual geographic distance between research partners could be measured, to permit appropriate weighting in the evaluation of university performance. Finally, while the research collaborations are measured in this model only in numeric terms, they could be also be evaluated qualitatively, in terms of impact. This could still be done by reference to bibliometric data, for example concerning the citations received by the joint publications, which represent the impact of the research results. A future line of research may embed the application to research institutes, in addition to universities The application of the model to the Italian case, presented here as an illustration, can be easily extended to other national situations, by applying the necessary specific adaptations. The model can also be sued to measure the intensity of international collaboration by universities with both public and private organizations.

The possible improvements described do not compromise one of the distinctive features of the instrument: the absolutely non-invasive character of its application, which avoids any necessity for any of the organizations under observation to furnish data. This guarantees, among other considerations, that the comparative evaluation is homogenous and reliable, with limited and definable costs in time and funds.